\documentstyle[12pt]{article}
\newcommand{\doublespace}{\renewcommand{\baselinestretch}{1.75}
\Large\normalsize}
\begin{document}
\doublespace
\begin{titlepage}

\centerline{\bf LOCALIZATION OF ENERGY IN GENERAL RELATIVITY}
\vskip 3.0cm
\centerline{Jos\'e W. Maluf$^*$}
\centerline{International Centre of Condensed Matter Physics}
\centerline{Universidade de Bras\'ilia}
\centerline{C.P. 04667}
\centerline{70.910-900 Bras\'ilia DF}
\centerline{Brazil}
\vskip 2.0cm

\date{}

\begin{abstract}
In the framework of the teleparallel equivalent of general relativity the
energy density $\varepsilon_g$  of asymptoticaly
flat graviational fields can be naturally  and unambiguously defined.
Upon integration of $\varepsilon_g$ over the whole three dimensional
space we obtain the ADM energy. We use $\varepsilon_g$ to calculate the
energy inside a Schawrzschild black hole.
\end{abstract}
\thispagestyle{empty}
\vfill
\noindent PACS numbers: 04.20.Cv, 04.20.Fy\par
\noindent (*) e-mail: wadih@fis.unb.br
\end{titlepage}

\newpage
\noindent The definition of localized energy density is a longstanding
problem in the theory of general relativity\cite{1}. The variation of the
Lagrangian density with respect to the metric field yields the
energy-momentum tensor of the standard field theories. It is well known
that this procedure cannot be satisfactorily applied to the
Hilbert-Einstein action. Usually it is asserted on the basis of the
principle of equivalence that the gravitational energy cannot be
localized\cite{MTW}. The principle of equivalence is frequently invoked to
assure that the gravitational field can be made to vanish in a sufficiently
small region of the spacetime: the dynamics of a system under the action
of a gravitational field in a locally inertial frame remains unchanged
if we eliminate the gravitational field but consider instead an appropriate
non-inertial frame. For this reason there have been attempts to define
{\it quasi-local} energy in general relativity (see ref.\cite{Brown} and
references therein).

The status of the principle of equivalence in the formulation of general
relativity, as a basic principle of the theory, is, to some extent,
controversial (see the discussion in the Preface of Synge's book\cite{Synge}).
{\it Einstein's} principle of equivalence\cite{Norton}
establishes an equivalence between inertial and non-inertial {\it reference
frames}, from what follows the equality between inertial and gravitational
masses, and ultimately the prescription according to which the gravitational
field couples in the same way with the other fields and matter systems
in nature (considering, of course, the different couplings between gravity
and boson and fermion fields). However, the symmetry group in general
relativity is the group of transformation of coordinates, and a coordinate
transformation can have no effect on the presence or absence of a
gravitational field. It is precisely the requirement of invariance under
general coordinate transformations that leads to the Einstein tensor.
Cartan\cite{Cartan} proved that the most general second rank tensor
$(i)$ constructed
in a coordinate independent way from the metric tensor and its first and
second partial derivatives, $(ii)$ having a vanishing divergence, and
$(iii)$ linear in the second derivatives is Einstein tensor
(Lovelock\cite{Lovelock} showed that one can dispense with $(iii)$). From
the mathematical point of view Einstein's principle of equivalence plays
no role in the determination of the gravitational field equations in
vacuo.

Therefore the principle of equivalence cannot be taken to rule out the
existence of an expression for the gravitational energy density. The absence
of the latter in the literature is really due to the description of the
gravitational field in terms of the Hilbert-Einstein action integral,
which is not an appropriate framework for such considerations.

In this paper we will show that in the context of the teleparallel
equivalent of general relativity (TEGR) there is a natural and simple
definition of energy density of the gravitational field for asymptoticaly
flat spacetimes. This expression is first noticed in the Hamiltonian
formulation of the TEGR; specifically, it arises in the integral form
of the Hamiltonian constraint. However, it also appears as a surface term
in the total Hamiltonian of the TEGR, and therefore it has a canonical
significance.

Notation: spacetime indices $\mu, \nu, ...$ and local Lorentz indices
$a, b, ...$ run from 0  to 3. In the 3+1 decomposition latin indices from
the middle of the alphabet indicate space indices, according to
$\mu \,=\,0, i$, $a\,=\,(0), (i)$. The tetrad field $e^a\,_\mu$ and the
(arbitrary) spin affine connection $\omega_{\mu ab}$ yield the usual
definitions of the torsion and curvature tensors: $R^a\,_{b\mu \nu}=
\partial_\mu \omega_\nu\,^a\,_b+\omega_\mu\,^a\,_c\omega_\nu\,^c\,_b\,-\,...$,
$T^a\,_{\mu \nu}=\partial_\mu e^a\,_\nu\,+\omega_\mu\,^a\,_b e^b\,_\nu
-\,...$. The flat spacetime
metric is fixed by $\eta_{(0)(0)}=-1$.

Let us briefly recall the TEGR\cite{Maluf}. The Lagrangian density in
empty spacetime is given by

$$L(e,\omega,\lambda)\;=\;-ke\Sigma^{abc}T_{abc}\,+\,
e\lambda^{ab\mu \nu}R_{ab\mu \nu}(\omega)\;.\eqno(1)$$

\noindent The constant $k$ equals $1\over {16\pi G}$, where $G$ is the
gravitational constant; $e\,=\,det(e^a\,_\mu)$ and $\lbrace
\lambda^{ab\mu \nu} \rbrace$ are Lagrange multipliers. The tensor
$\Sigma^{abc}$ is defined as

$$\Sigma^{abc}\;=\;{1\over 4}(T^{abc}\,+\,T^{bac}\,-\,T^{cab})\,+\,
{1\over 2}(\eta^{ac}T^b\,-\,\eta^{ab}T^c)\;,$$

\noindent which yields

$$\Sigma^{abc}T_{abc}\;=\;{1\over 4}T^{abc}T_{abc}\,+\,{1\over 2}
T^{abc}T_{bac}\,-\,T^aT_a\;,$$

\noindent where $T_a\,=\,T^b\,_{ba}$.

The affine connection $\omega_{\mu ab}$ can be (identically) rewritten
as $\omega_{\mu ab}\,=\,^o\omega_{\mu ab}\,+\,K_{\mu ab}$, where
$^o\omega_{\mu ab}$ is the Levi-Civita connection and
$K_{\mu ab} \,=\,{1\over 2}e_a\,^\lambda e_b\,^\nu(T_{\lambda \mu \nu}+
T_{\nu \lambda \mu}-T_{\mu \nu \lambda})$ is the contortion tensor. Such
decomposition of $\omega_{\mu ab}$ allows us to obtain the identity

$$eR(e,\omega)\;=\;eR(e)\,+\,e\Sigma^{\alpha \mu \nu}T_{\alpha \mu \nu}\;-\;
2\partial_\mu(eT^\mu)\;,\eqno(2)$$

\noindent by just substituting $\omega_{\mu ab}$ for
$^o\omega_{\mu ab}+K_{\mu ab}$ on the left hand side (LHS) of (2). We
observe that the vanishing of $R^a\,_{b\mu \nu}(\omega)$ implies the
equivalence of the scalar curvature $R(e)$ - constructed out of $e^a\,_\mu$
only - and the quadratic combination of the torsion tensor (for
asymptoticaly flat spacetimes the divergence in (2) does not contribute
to the action integral).

Except for the presence of $-k$ in (1), the Lagrangian density considered
here is the same as in ref.\cite{Maluf}. Variation of $L$ with respect to
$\lambda^{ab\mu \nu}$ yields

$$R^a\,_{b\mu \nu}(\omega)\;=\;0\;.\eqno(3)$$

\noindent Let ${{\delta L}\over {\delta e^{a\mu}}}=0$ denote the field
equation satisfied by $e^a\,_\mu$. With the help of (3) it can be shown
by explicit calculations that

$${{\delta L}\over{\delta e^{a\mu}}}\;=\;{1\over 2}
\lbrace R_{a\mu}(e)\,-\,{1\over 2}e_{a\mu}R(e)\rbrace\;,\eqno(4)$$

\noindent again by replacing $\omega_{\mu ab}$ for $^o\omega_{\mu ab}(e)+
K_{\mu ab}$ in the LHS of (4). Therefore $e^a\,_\mu$ does satisfy
Einstein's equations.

The field equations arising from variations of $L$ with respect to
$\omega_{\mu ab}$ can be best analysed in the Hamiltonian formulation.
The latter has been presented in ref.\cite{Maluf}, with the gauge
$\omega_{0ab}=0$ being fixed from the outset. In this paper we will
likewise maintain this gauge fixing, as it can be shown that in this
restricted context the constraints of the theory constitute a first class
set. The condition $\omega_{0ab}=0$ is fixed by breaking the local Lorentz
symmetry of (1). We still make use of the residual time independent
gauge symmetry to fix the usual time gauge condition
$e_{(k)}\,^0\,=\,e_{(0)i}\,=\,0$.

The Hamiltonian density $H$ can be successfully constructed out of (1) in
terms of canonical field variables and Lagrange multipliers.
Because of the gauge fixing $\omega_{0ab}=0$, $H$
does not depend on $P^{kab}$, the momentum
canonically conjugated to $\omega_{kab}$. Thus arbitrary variations
of $L\,=\,p\dot q -H$ with respect to $P^{kab}$ yields
$\dot{\omega}_{kab}\,=\,0$ (see expression (21) of the Lagrangian density
in [8]; had we not fixed $\omega_{0ab}=0$ the corresponding equation
would be $\dot{\omega}_{kab}\,-\,D_k\omega_{0ab}\,=\,0$, which is equivalent
to the Lagrangian field equation $R_{ab 0k}=0$). Therefore in view of
$\omega_{0ab}=0$, $\omega_{kab}$ drops out from our considerations. The
above gauge fixing can be understood as the fixation of a {\it global}
reference frame.

The details of the 3+1 decomposition of (1) are given in [8], except that
the surface terms in eq. (6) below were not considered in the latter.
The canonical action integral becomes

$$A_{TL}\;=\;\int d^4x\,\lbrace \Pi^{(j)k}\dot e_{(j)k}\,-\,H\rbrace\;,
\eqno(5)$$

$$H\;=\;NC\,+N^iC_i\,+\Sigma_{mn}\Pi^{\lbrack mn \rbrack}\,+\,
{1\over 8\pi G}\partial_k(NeT^k)\,+\,\partial_k(\Pi^{jk}N_j)\;.\eqno(6)$$

\noindent $N$ and $N^i$ are the lapse and shift functions, and
$\Sigma_{mn}\,=\,-\Sigma_{nm}$
are Lagrange multipliers. The constraints are defined by

$$C\;=\;\partial_j(2keT^j)\,-\,ke\Sigma^{kij}T_{kij}\,-\,
{1\over {4ke}}(\Pi^{ij}\Pi_{ji}-{1\over 2}\Pi^2)\;,\eqno(7)$$

$$C_k\;=\;-e_{(j)k}\partial_i\Pi^{(j)i}\,-\,\Pi^{(j)i}T_{(j)ik}\;,\eqno(8)$$

\noindent with $e=det(e_{(j)k}$ and $T^i=g^{ik}e^{(j)l}T_{(j)lk}$. We
remark that (5) is invariant under global SO(3) and general coordinate
transformations. An important feature of this framework is that although
we are considering asymptoticaly flat gravitational fields, the
action integral determined by (1) does {\it not} require any additional
surface term, as it is invariant under coordinate transformations that
preserve the asymptotic structure of the field quantities. A clear discussion
concerning the necessity of the addition of a surface term to the
Hilbert-Einstein action $A_{HE}$, in the case of asymptotically flat
gravitational fields, is given by Faddeev\cite{Faddeev}.

We consider now that for $r \rightarrow \infty$ we have
$e_{(j)k} \approx \eta_{jk}+{1\over 2}h_{jk}({1\over r})$.
In view of the relation

$${1\over {8\pi G}}\int d^3x\,\partial_j(eT^j)\;=\;
{1\over {16\pi G}}\int_S dS_k(\partial_i h_{ik}-\partial_k h_{ii})\;
\equiv \;E_{ADM}\;,\eqno(9)$$

\noindent where the surface integral is evaluated for $r\rightarrow \infty$,
we observe that the integral form of the Hamiltonian constraint
$C=0$ may be rewritten as

$$\int d^3x \biggl\{ ke\Sigma^{kij}T_{kij}\,+\,
{1\over {4ke}}(\Pi^{ij}\Pi_{ji}-{1\over 2}\Pi^2) \biggr\}\;=\;
E_{ADM}\;;\eqno(10)$$

\noindent the integration is over the whole tree dimensional space. Since
$\partial_j(eT^j)$ is a scalar density, from (9) and (10) we are naturally
led to define the gravitational energy enclosed by a volume $V$ of the
space as

$$E_g\;=\;{1\over {8\pi G}}\int_V d^3x\partial_j(eT^j)\;.\eqno(11)$$

\noindent The expression above is manifestly invariant under general
coordinate transformations of the three dimensional space-like
hypersurface $\Sigma$ and yields $E_{ADM}$ if the integration is over
the whole $\Sigma$.

Let us note from (6) that $\int d^3x H$ evaluated from a set
$(e_{(j)k}, \Pi^{(k)l})$ that satisfy the field equations, in a coordinate
system such that for $r\rightarrow \infty$ we have $N=1,\,N_j=0$, also
yields $E_{ADM}$.

The appearance of a scalar and vector densities in (6) is intimately related
to the fact that it is not necessary to add a non-covariant surface term
to (1). The surface term that must be added to the Hilbert-Einstein action
is not well behaved under coordinate transformations, but as Faddeev
stressed, the action integral is invariant\cite{Faddeev}. Since it is
precisely this surface term that leads to $E_{ADM}$, it is not possible
to obtain a satisfactory energy density out of $A_{HE}$.

A similar difficulty occurs in the context of the Einstein-Cartan
formulation. The Lagrangian density for asymptotically flat gravitational
fields in the first order differential formulation is given by

$$L_{EC}\;=\;ee^{a\mu}e^{b\nu}R_{ab\mu \nu}(\omega)\;-\;
2\partial_\mu (ee^{a\mu}e^{b\nu}\omega_{\nu ab})\;.\eqno(12)$$

\noindent After performing a canonical 3+1 decomposition of $L_{EC}$,
the Hamiltonian density acquires a surface term which reads\cite{Ma2}

$$\varepsilon_{{}_{EC}}\;=\;\partial_i(2Nee^{ai}e^{bj}\,^o\omega_{jab}\,-\,
N^je^{ai}\Pi_{aj})\;.\eqno(12)$$

\noindent In the appropriate limit $\int d^3x \varepsilon_{{}_{EC}}$ yields
$E_{ADM}$, but no energy density can be defined from (11).

The gravitational energy-momentum four-vector of theories with local gauge
symmetry has already been discussed in the literature, in the context
of Poincar\'e gauge theories\cite{Hehl}.
Such analyses are always carried out in
the Lagrangian framework, and can equally well be applied to the TEGR.
We note, however, that the present considerations are derived from the
Hamiltonian formulation. To our knowledge, the appearance of the ADM
energy in the Hamiltonian constraint has not been previously noticed.

Brown and York\cite{Brown} have recently furnished an expression of
quasilocal gravitational energy density for compact geometries, by
resorting to a Hamilton-Jacobi-type analysis. They consider the action
integral evaluated from field quantities that solve the classical
equations of motion, and as a function of the time interval. Then the
action $A$ satisfies the Hamilton-Jacobi equation
$E\,=\,-{{\partial A}\over {\partial T}}$, where $E$ expresses the energy
of the classical solution. In analogy with this procedure they define
the quasilocal energy associated with a space-like hypersurface $\Sigma$
as minus the variation of the action with respect to a unit increase in
proper time separation between the boundary $B$ of $\Sigma$ and its
neighboring two-surface, as measured orthogonally to $\Sigma$ at $B$.
In the present case we can similarly define the energy density as minus
the variation of the action with respect to the proper time $N(x)$. For
a given set of solutions of the classical equations of motion the energy
density $\varepsilon_g$ can be defined as

$$\varepsilon_g(x)\;=\;-{{\delta A_{TL}}\over {\delta N(x)}}\;=\;
{1\over{8\pi G}}\partial_j(eT^j)\;,\eqno(14)$$

\noindent in agreement with (11).

The fixation of the time gauge condition prevents the construction of
a global SO(3,1) energy-momentum four vector $P^a$, because the space
and time components of $P^a$ can no longer be mixed. Nevertheless we
can identify the components of such a vector. The field quantities
$\lbrace \Pi^{ak}\rbrace$ are defined by $\Pi^{ak}\,=\,
{{\delta L}\over {\delta \dot e_{ak}}}$. $\Pi^{(j)k}$ is the true
momentum canonically conjugated to $e_{(j)k}$. However, because of the
time gauge condition, in the 3+1 decomposition $\Pi^{(0)k}$ must be
expressed in terms of canonical field variables. One ultimately finds
$\Pi^{(0)k}\,=\,2eT^k$ (see the discussion following eq. (19) of [8]).

We assume now that $e_{(j)k}$ and $\Pi^{(j)k}$ satisfy the constraints.
{}From $C_i=0$ we obtain

$$\partial_i\Pi^{(m)i}\;=\;-e^{(m)k}\Pi^{(j)i}T_{(j)ik}\;.\eqno(15)$$

\noindent Therefore we define the energy-momentum quadruplet $P^a$ as

$$P^a\;=\;{1\over {16\pi G}}\partial_k\Pi^{ak}\;.\eqno(16)$$

\noindent Since the constraints are satisfied we have

$$P^{(0)}\;=\;{1\over {16\pi G}}\partial_j(eT^j)\;=\;
{1\over {4ke}}(\Pi^{ij}\Pi_{ji}-{1\over 2}\Pi^2)\,+\,
ke\Sigma^{kij}T_{kij}\;,\eqno(17)$$

$$P^{(m)}\;=\;-{1\over {16\pi G}}e^{(m)k}\Pi^{(j)i}T_{(j)ik}\;.\eqno(18)$$

In the following we will specialize the Hamiltonian formulation to the
spherically symmetric geometry, in order to compute the gravitational
energy inside a Schwarzschild black hole. We fix the triads
$e_{(k)i}$ as

$$e_{(k)i}\;=\;\pmatrix{ e^\lambda\,sin\theta cos\phi&r\,cos\theta cos\phi
&-r\, sin\theta sin\phi \cr
e^\lambda\,sin\theta sin\phi & r\,cos\theta sin\phi
& r\, sin\theta cos\phi \cr
e^\lambda\,cos\theta & -r\,sin\theta &0\cr }\eqno(19)$$

\noindent where $\lambda\,=\,\lambda(r,t)$; $(k)$ is the line index and
$i$ is the column index. The one form $e^{(k)}$ is defined as

$$e^{(k)}\;=\;e^{(k)}\,_r dr\,+\,e^{(k)}\,_\theta d\theta\,+\,
e^{(k)}\,_\phi d\phi\;,$$

\noindent from what follows

$$e^{(k)}e_{(k)}\;=\;e^{2\lambda}dr^2\,+\,r^2\,d\theta^2\,+\,
r^2\,sin^2 \theta\, d\phi^2\;.$$

\noindent We also obtain $e\,=\,det(e_{(k)i})\,=\,r^2\,sin\theta\,e^\lambda$.
For $r \rightarrow \infty$ we require $\lambda(r) \sim O({1\over r})$.

The symmetry reduction is performed directly in the Hamiltonian. We first
determine the Killing vectors $\xi$ of $g_{ij}=e^{(k)}\,_i e_{(k)j}$.
Next we require the vanishing of the Lie derivative
$L_\xi(e^{-1}\Pi_{ij})=0$, where $\Pi_{ij}=g_{jk}e_{(l)i}\Pi^{(l)k}$. We
obtain

$$e^{-1}\Pi_{ij}\;=\;diag(A(r,t), B(r,t), B(r,t)sen^2\theta)\;;\eqno(20)$$

\noindent $A(r,t)$ and $B(r,t)$ are arbitrary functions. From (20) we can
calculate all $\lbrace \Pi^{(k)j} \rbrace$. Upon substitution of the latter
and (19) into (5) we find out, as expected,  that there is no canonical
field quantity conjugated to $B(r,t)$. Thus we enforce $B(r,t)=0$,
which implies in $\Pi^{(k)2}\,=\,\Pi^{(k)3}\,=0$. Defining $\Pi$ by
$\Pi\,=\,kr^2\,e^{-\lambda}\,A$ and integrating over angles we finally
obtain the action integral

$$A\;=\;4\pi \int dt\,dr\,\lbrace \Pi\dot{\lambda}\,-\,NC\,-\,
N^1C_1\rbrace\;,\eqno(21)$$

$$C\;=\;2ke^\lambda(1-e^{-\lambda})^2\,-\,{1\over {8kr^2}}e^{-\lambda}\Pi^2
\,+\,{1\over {4\pi}}\varepsilon\;,\eqno(22)$$

$$C_1\;=\;-e^\lambda{\partial \over {\partial r}}
(e^{-\lambda}\Pi)\;,\eqno(23)$$

$$\varepsilon\;=\;16\pi k{\partial \over {\partial r}}
\lbrack r(1-e^{-\lambda})\rbrack\;,\eqno(24)$$

\noindent The constraints $C_2=C_\theta$ and $C_3=C_\phi$ vanish identically.

The Hamiltonian formulation established by (21-24) is completely equivalent
to the corresponding construction in the framework of the ADM
formulation\cite{Berger}, as it can be shown that the Hamiltonian and
vector constraints are equivalent in both cases. If we choose a coordinate
system such that $N^1=0$, then the constraints and the evolution equations
for $\lambda$ and $\Pi$ yield the Schwarzschild solution,

$$e^{-2\lambda}\;=\;1\,-\,{2mG\over r}\;,\eqno(25)$$

\noindent together with $N^2=e^{-2\lambda}$.

The total energy associated with (25) may be calculated from the surface
term (24). In analogy with (9) we obtain, as expected,

$$E_{ADM}\;=\;\int^\infty dr\,\varepsilon_g(r)\;=\;m\;.\eqno(26)$$

\noindent We can also compute exactly the gravitational energy  inside
a black hole. It is given by

$$E_{BH}\;=\; \int _0^{2mG} dr\,\varepsilon_g(r) \;=\;2m\;,\eqno(27)$$

\noindent which implies that the total energy exterior to the black hole
is $-m$.

\noindent It is of interest to evaluate, in addition, the
the gravitational energy inside a spherical surface of arbitrary
radius $R$:

$$E_g\;=\;R \biggl\{ 1\,-\,(1-{{2m}\over R})^{1\over 2}
\biggr\}\;.\eqno(28)$$

\noindent This is exactly the expression found by Brown and
York\cite{Brown} in their analysis of {\it quasilocal} gravitational
energy of the Schwarzschild solution. The method developed by
Brown and York, however, does not seem to be applicable to an
arbitrary metric field. Problems appear in the calculation of the
quasilocal energy in the framework of the Kerr metric\cite{Martinez}.
On the contrary, given the triad components restricted to a three
dimensional hypersurface of the Kerr type we can easily calculate
$E_g$ by means of (14).

In conclusion, we find out that in the TEGR there is a natural, consistent
and unambiguous definition of gravitational energy density. This fact
indicates that this framework is suitable for the Hamiltonian analysis
of the gravitational field. In particular, the integral form of the
Hamiltonian constraint, eq.(10), may become an energy eigenvalue equation
in the canonical quantization program.

\vskip 2.0cm
\noindent {\it Acknowledgement}\par
\noindent This work was supported in part by CNPQ.\par

\end{document}